\documentclass[3p,times,procedia]{elsarticle}
\usepackage{nupha_ecrc}

\volume{00}

\firstpage{1}

\journalname{Nuclear Physics A}

\runauth{}

\jid{nupha}

\jnltitlelogo{Nuclear Physics A}

\usepackage{mathrsfs}
\usepackage{amssymb}
\usepackage{amsmath}
\usepackage{graphicx}
\usepackage[utf8]{inputenc}
\usepackage{amsfonts,amsbsy}
\usepackage{nicefrac}

\usepackage[figuresright]{rotating}

\newcommand{\mrm}{\mathrm}

\newcommand{\fig}{Fig.~}

\newcommand{\bs}[1]{\boldsymbol{#1}}

\newcommand{\ud}{\mathrm{d}}

\begin{document}

\begin{frontmatter}



\dochead{XXVIIIth International Conference on Ultrarelativistic Nucleus-Nucleus Collisions\\ (Quark Matter 2019)}

\title{Heavy quark momentum diffusion coefficient in 3D gluon plasma}


\author{K. Boguslavski$^1$, A. Kurkela$^{2,3}$, T. Lappi$^{4,5}$, J. Peuron$^6$}

\address{$^1$ Institute  for  Theoretical  Physics,  Technische  Universität  Wien,  1040  Vienna,  Austria \\ $^2$ Department  of  Physics,  University  of  Jyvaskyla,P.O.  Box  35,  40014  University  of  Jyväskylä,  Finland \\ $^3$ Theoretical  Physics  Department,  CERN,  Geneva,  Switzerland \\ $^4$ Faculty  of  Science  and  Technology,  University  of  Stavanger,  4036  Stavanger,  Norway \\ $^5$ Helsinki  Institute  of  Physics,  P.O.  Box  64,  00014  University  of  Helsinki,  Finland \\ $^6$ European Centre for Theoretical Studies in Nuclear Physics and Related Areas(ECT*) and Fondazione Bruno Kessler ,Strada  delle  Tabarelle  286,  I-38123  Villazzano  (TN),  Italy}

\begin{abstract}
We study the heavy-quark momentum diffusion coefficient in far from equilibrium gluon plasma in a self-similar regime using real-time lattice techniques. We use 3 methods for the extraction: an unequal time electric field 2-point correlator integrated over the time difference, a spectral reconstruction (SR) method based on the measured equal time electric field correlator and a kinetic theory (KT) formula. The time-evolution of the momentum diffusion coefficient extracted using all methods is consistent with an approximate $t^{\nicefrac{-1}{2}}$ power law. We also study the extracted diffusion coefficient as a function of the upper limit of the time integration and observe that including the infrared enhancement of the equal-time correlation function in the SR calculation improves the agreement with the data for transient time behavior considerably. This is a gauge invariant confirmation of the infrared enhancement previously observed only in gauge fixed correlation functions.
\end{abstract}


\end{frontmatter}


\section{Introduction}
\label{sec:intro}
Evaluation of the transport properties of the QGP created in collider experiments such as the LHC and RHIC has been a longstanding puzzle for the heavy-ion theory community. However, in conventional transport calculations the effects of the pre-equilibrium phase are usually completely ignored. Only very recently the transport properties of the glasma have been addressed using Fokker-Planck transport equations \cite{Mrowczynski:2017kso,Carrington:2020sww} and also classical real time lattice simulations \cite{Ipp:2020mjc}. It also seems that the glasma phase can substantially contribute to jet quenching  \cite{Carrington:2020sww}. Also the later stages of the non-equilibrium evolution have been studied using a kinetic theory approach  \cite{Das:2017dsh,Song:2019cqz}. These studies have indicated that the pre-equilibrium effects can be important. Our aim in this paper is to study the heavy quark momentum diffusion coefficient in a far from equilibrium gluon plasma in the self-similar regime using classical gluodynamics. 

\section{Evaluation of the correlation function, 3 methods}
\begin{figure}
\includegraphics[scale=0.5]{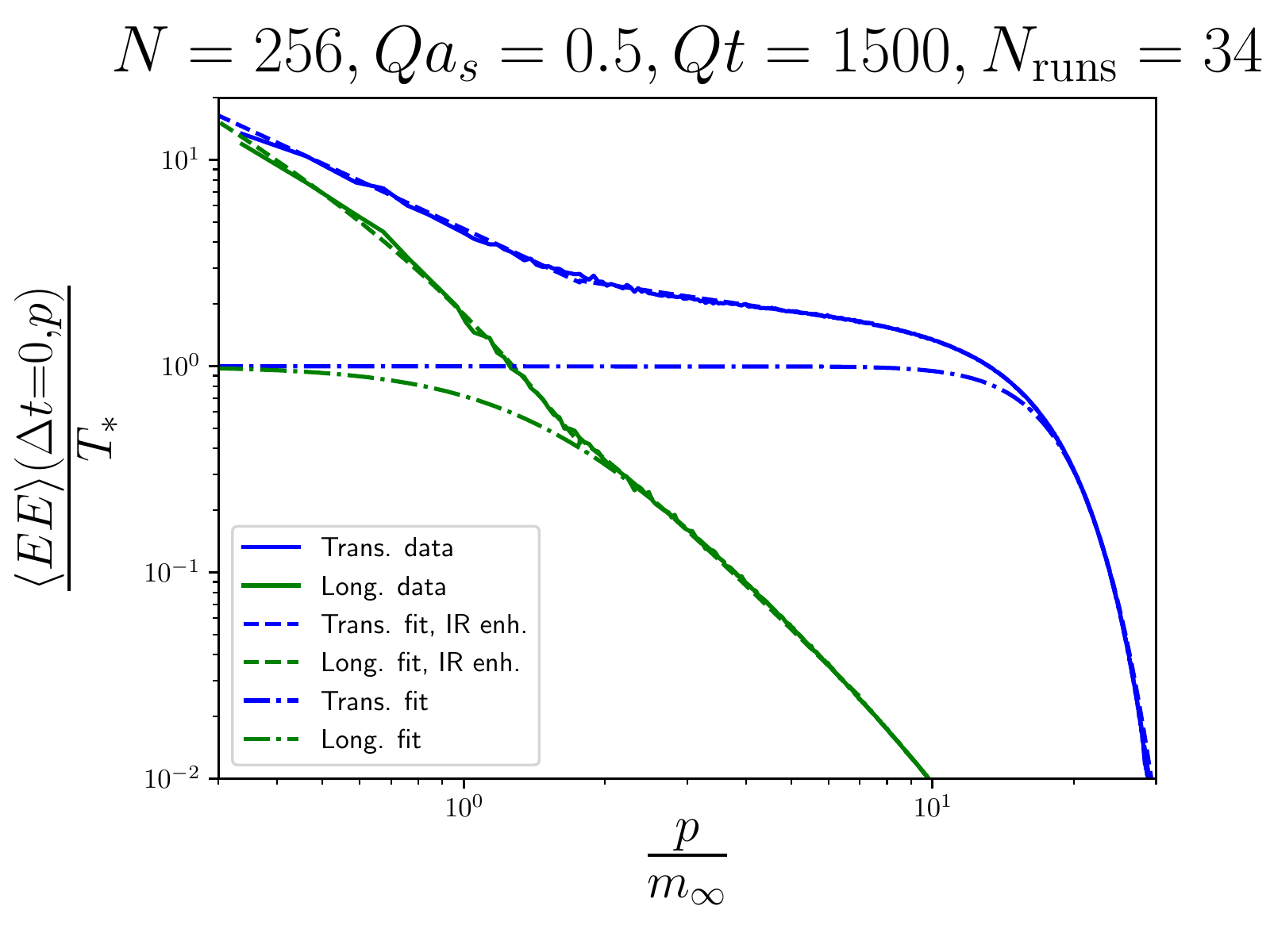}
\includegraphics[scale=0.5]{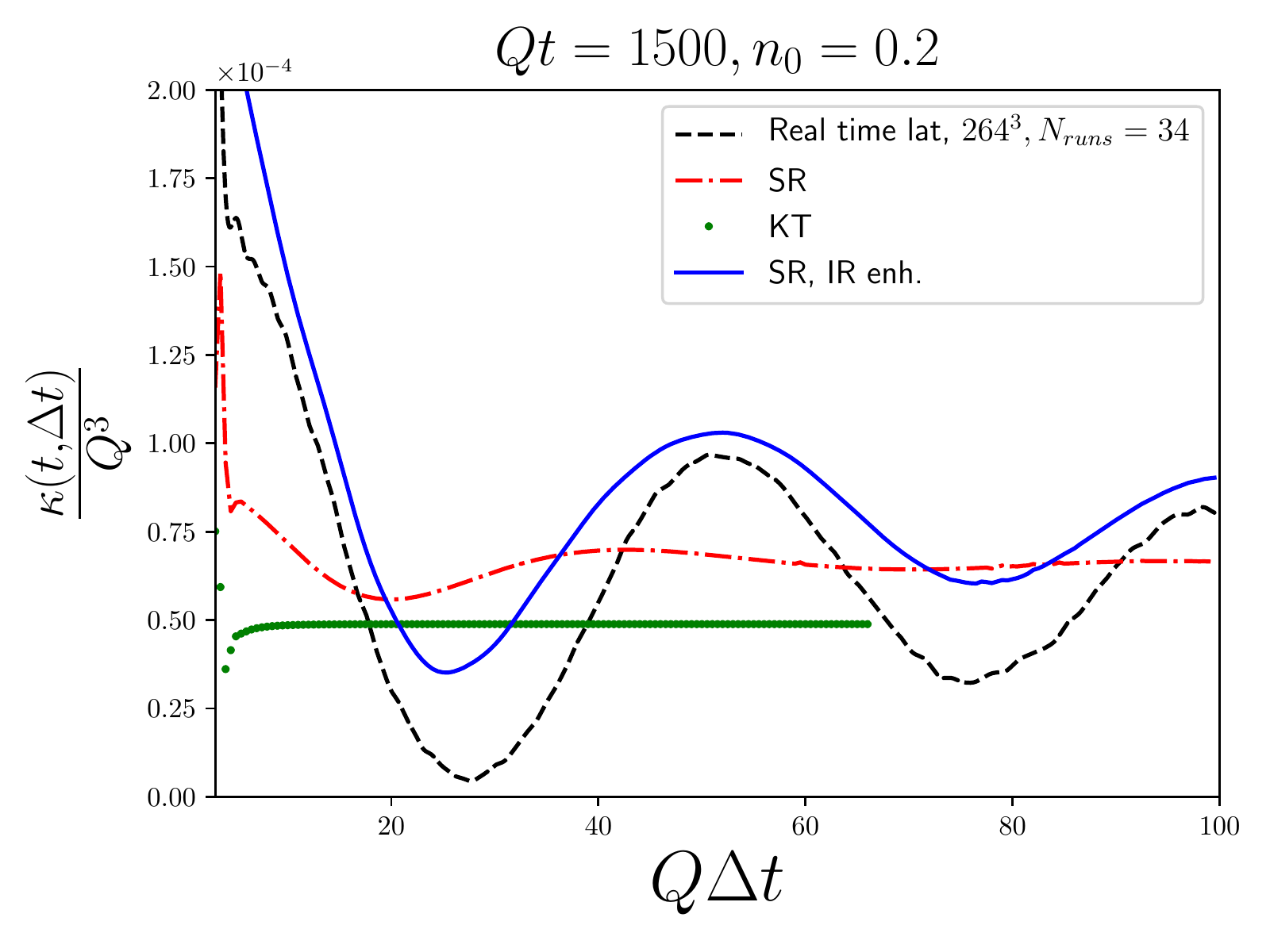}
\caption{(Left) Extracted equal time transverse and longitudinal statistical functions. The fit  of the data is shown using dashed lines, corresponding to the infrared enhanced equal time statistical correlation function. The dash dotted lines correspond to the unenhanced correlator. The HTL expectation is then smoothly matched to data to incorporate proper the UV behavior. (Right) Extracted value of $\kappa\left(Q t = 1500 , \Delta t \right)$ as a function of the upper time integration limit $\Delta t$. The signal oscillates at the plasma frequency. The input to the SR calculation (the mass scales etc.) have been matched to their values. }
\label{fig:statfun}
\end{figure}

\subsection{Real time lattice measurement}
In temporal gauge the heavy quark momentum diffusion coefficient can be estimated as \cite{CasalderreySolana:2006rq,CaronHuot:2009uh}
 \begin{align}
\label{eq:kappa_from_data}
\kappa\left(t , \Delta t \right) &=  \dfrac{1}{3 N_c} \int_{t}^{t + \Delta t} \mathrm{d} t^\prime  \dfrac{1}{V} \int \mathrm{d}^3 x \sum_{i = 1}^3 \left< g E_i^a \left(\boldsymbol{x} ,t \right) g E_i^a  \left(\boldsymbol{x}, t^\prime \right)\right>.
\end{align}
 In this approximation we assume that the heavy quark is infinitely heavy and the kicks it gets from the medium are due to the (nonabelian) Lorentz force. This leads to momentum broadening $\frac{d}{dt}\langle p^2 \rangle = 3\kappa_\infty(t)$, where $\kappa_\infty(t) \equiv \lim_{\Delta t \rightarrow \infty} \kappa(t,\Delta t)$ is the heavy-quark diffusion coefficient often used for effective equations of heavy quarks and quarkonia \cite{CasalderreySolana:2006rq,CaronHuot:2009uh,Brambilla:2016wgg}.

\subsection{Spectral reconstruction method (SR) and infrared enhancement}
Our starting point is \eqref{eq:kappa_from_data}. The correlation function corresponds to the unequal time statistical correlation function. This can be expressed  using the  generalized fluctuation dissipation relation 
\begin{align}
 \dfrac{\langle EE \rangle_{T/L}(t, \omega , p)}{\langle EE \rangle_{T/L}(t, \Delta t=0 , p)} = \dfrac{\dot{\rho}_{T/L}(t, \omega , p)}{\dot{\rho}_{T/L}(t, \Delta t=0 , p)},
\end{align}
where $\langle EE \rangle$ is the statistical correlation function and $\dot{\rho}$ is the spectral function. The equal time statistical correlation functions are shown in \fig \ref{fig:statfun} along with two possible parametrizations.
Previously we have observed in \cite{Boguslavski:2018beu} an infrared modes are enhanced in the statistical correlator when compared to HTL expectations. One parametrization matches the HTL expectation in the infrared and  to our data in the UV (dash-dotted lines). The other parametrization uses our data over the whole spectrum (dashed lines). 
The spectral function $\rho$ consists of quasiparticle and Landau damping contributions. Our final expression for the diffusion coefficient is $\kappa_{\text{SR}}(t,\Delta t) = \kappa_{\text{SR}}^{\text{Landau}}(t,\Delta t) + \kappa_{\text{SR}}^{\text{Particles}}(t,\Delta t)$ with
\begin{align}
\label{eq:LandauContribution}
\kappa^{\mrm{Landau}}_{\text{SR}}\left( t, \Delta t \right) =&\, \dfrac{d_A}{6 N_c} \int \dfrac{\mathrm{d}^3p}{\left( 2 \pi \right)^3} \int_{-1}^{1} \ud x\; 2\,p \sin{\left(x\,p \Delta t\right)} \Bigg[2\, \dfrac{g^2 \langle EE \rangle_T(t,\Delta t=0,p)}{\dot{\rho}_{T}(t, \Delta t=0 , p)}\,\beta_T(t, \omega , p) \nonumber  \\  &  \qquad  + \dfrac{g^2 \langle EE \rangle_L(t,\Delta t=0,p)}{\dot{\rho}_{L}(t, \Delta t=0 , p)}\,\dfrac{\beta_L(t, \omega , p)}{x^2} \Bigg] \\
\label{eq:ParticleContribution}
\kappa^{\mrm{Particles}}_{\text{SR}}\left( t, \Delta t \right) =&\, \dfrac{d_A}{6 N_c} \int \dfrac{\mathrm{d}^3p}{\left( 2 \pi \right)^3}\, \left[2\, \dfrac{g^2 \langle EE \rangle_T(t,\Delta t=0,p)}{\dot{\rho}_{T}(t, \Delta t=0 , p)}\,4\, Z_{T}(p) \sin{\left(\omega_{T}(p) \Delta t \right)}\, e^{-\gamma_{T}(p) T} \right. \nonumber \\
&\, \qquad \qquad \quad \left.+ \dfrac{g^2 \langle EE \rangle_L(t,\Delta t=0,p)}{\dot{\rho}_{L}(t, \Delta t=0 , p)}\,4\,\dfrac{p^2 Z_L(p)}{\omega_{L}(p)^2}\,\sin{\left(\omega_{L}(p) \Delta t \right)}\, e^{-\gamma_{L}(p) T} \right],
\end{align}
where $t$ is the starting time of the measurement, $\Delta t$ is the size of the time-integration window, $\beta_{T,L}$ is the transverse (longitudinal) Landau damping contribution and $Z_{T,L}$ are the residues of the transverse (longitudinal) quasiparticle peaks, $\omega_{T,L}$ is the transverse (longitudinal) dispersion relation, $\gamma$ is the quasiparticle damping rate and $x = \nicefrac{\omega}{p}$. The expressions listed in this section can be found in e.g.  \cite{Boguslavski:2018beu,lebellac_1996}, and we have numerically extracted them for pure glue systems in \cite{Boguslavski:2018beu} using the linear response framework outlined in \cite{Kurkela:2016mhu}.

\subsection{Kinetic theory}
Next we estimate $\kappa\left(t\right)$ in a kinetic theory framework following closely \cite{Moore:2004tg}. In order to take finite time effects into account we allow  a nonzero frequency $\omega = k^\prime - k$ corresponding to finite energy transfer between the gluon and quark. In this case the matrix element becomes 
\begin{align}
 \left|\mathcal{M} \right|^2_{\mathrm{gluon}}(\omega) = \left[N_c C_H g^4 \right] \dfrac{4 M^2 (k_0 + k_0')^2 \left( 1 + \cos^2 \theta_{\bs{k}\bs{k}'} \right)}{(q^2-\omega^2+m_D^2)^2}.
\end{align}
The frequency dependent expression can be written as 
\begin{align}
\label{eq:KT_omGen}
\kappa(\omega) \,&= \dfrac{1}{6 M} \int \dfrac{\mathrm{d}^3 \boldsymbol{k}  \mathrm{d}^3 \boldsymbol{q}}{\left(2 \pi\right)^6 8 |\bs{k}| |\bs{k+q}| M} 2 \pi \delta(|\bs{k + q}| - |\bs{k}| - \omega)\, \boldsymbol{q}^2 \left|\mathcal{M} \right|^2_{\mathrm{gluon}}(\omega) f(k)f(\left| \bs{k} + \bs{q} \right|).
\end{align}
This integral is evaluated numerically using the distribution at $Qt=1500$. 
 
\section{Dependence on time and occupation number}
\begin{figure}
\includegraphics[scale=0.41]{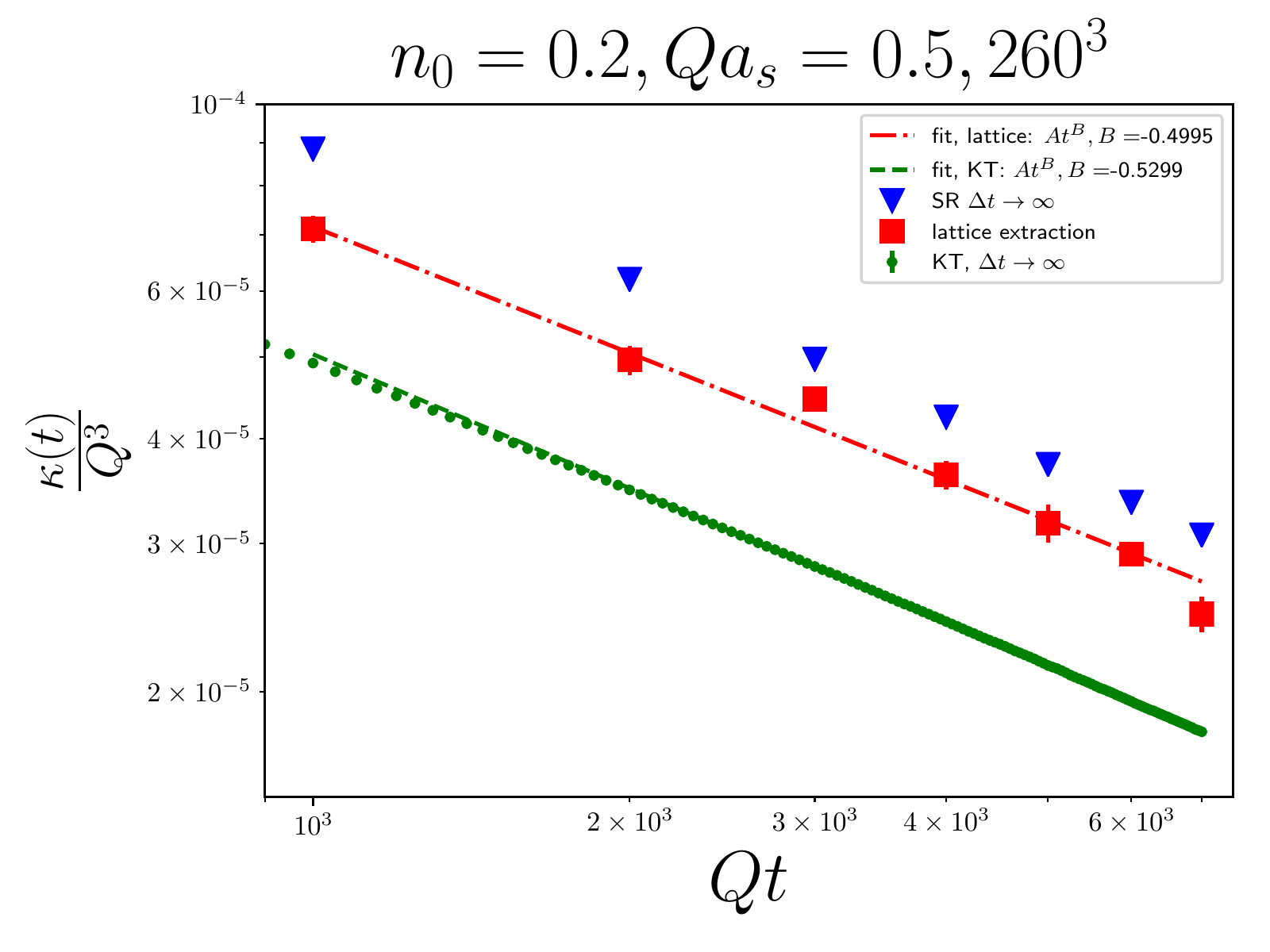}
\includegraphics[scale=0.41]{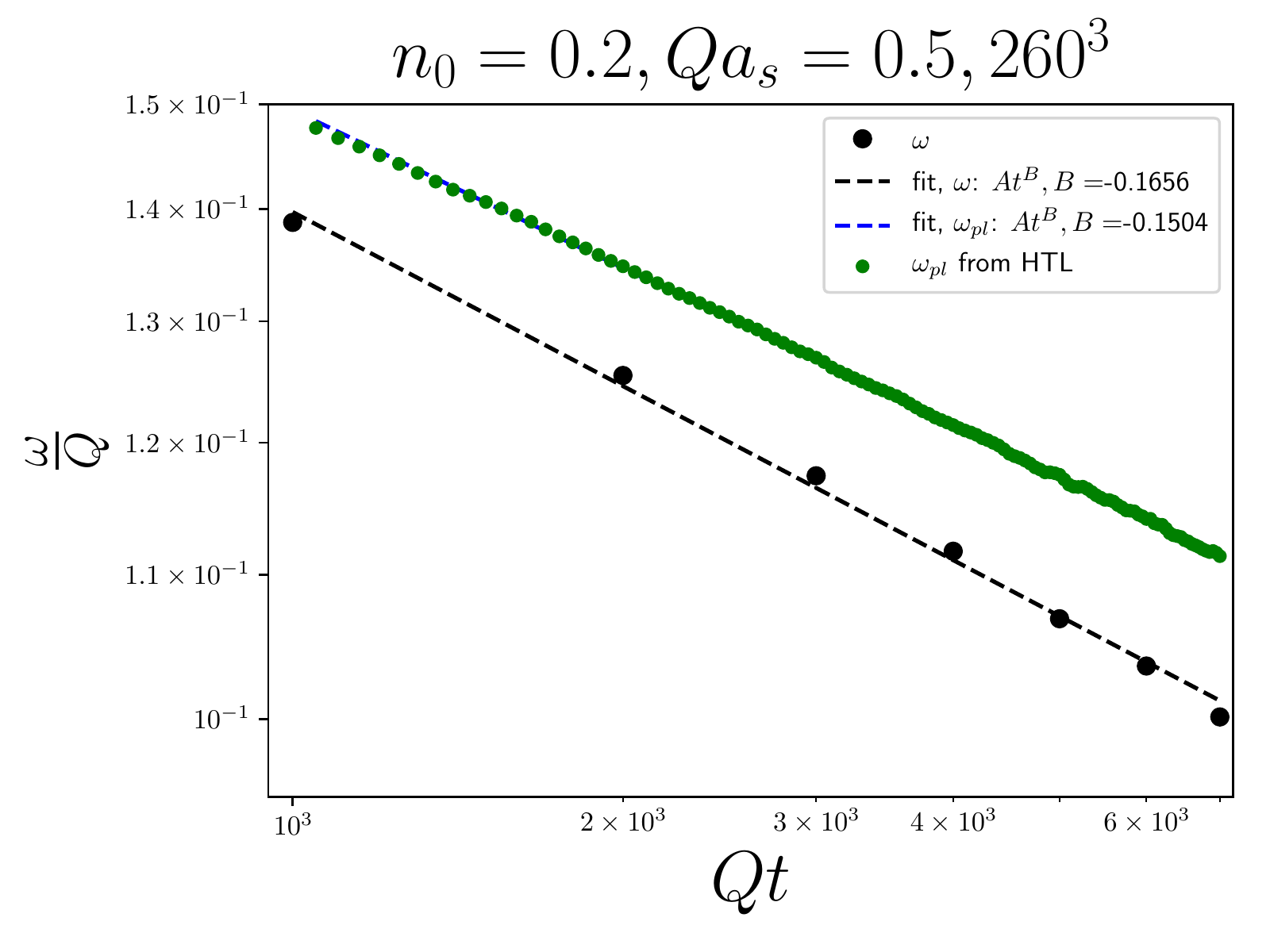}
\caption{(Left) Dependence on time $t$, averaged over 15 runs. We also show the extracted value using kinetic theory and SR  at infinite time limit. (Right) Extraction of plasmon frequency  and the extracted oscillation frequency of $\kappa(t , \Delta t)$ in $\Delta t $ as a function of $t$.}
\label{fig:time_dep}
\end{figure}

The dependence on the time integration window size $\Delta t $ is shown in \fig \ref{fig:statfun} (right). The curve extracted from real time lattice simulations features an oscillation roughly at the plasmon frequency, adding evidence to an enhancement of low-energy gluonic excitations \cite{Berges:2019oun}. The SR method without the infrared enhancement does not reproduce these oscillations, and neither does kinetic theory. However when the infrared enhancement is taken into account we observe oscillations with a similar frequency. The main contributions to the IR enhanced SR curve are the quasiparticle contributions, which give rise to the oscillation, and  the longitudinal Landau damping, is responsible for the value at $\Delta t \to \infty$. \fig \ref{fig:time_dep} (right) demonstrates that the oscillation frequency in $\Delta t$ corresponds to the plasmon frequency for various $t$ following the approximate $t^{\nicefrac{-1}{7}}$ powerlaw of the plasmon frequency in the self-similar regime \cite{Lappi:2016ato}.

The dependence on the time of the measurement t is shown in  \fig \ref{fig:time_dep} (left). We observe that $\kappa$ follows an approximate $t^{\nicefrac{-1}{2}}$ powerlaw.

\section{Summary}
We have measured the heavy quark diffusion coefficient in classical gluodynamics using three different methods, which are in good agreement. The first method was direct lattice measurement of the appropriate correlation function. The second method is spectral reconstruction using HTL perturbation theory and our data on the spectral properties of the gluons. The third method is a kinetic theory calculation with the lattice extracted occupation number. The infrared enhancement of the statistical correlation function introduces oscillations in $\kappa(t , \Delta t)$ in $\Delta t$ with the plasmon frequency. These oscillations are not reproduced by kinetic theory. The time dependence of the diffusion coefficient follows an approximate $t^{\nicefrac{-1}{2}}$ powerlaw.

\section*{Acknowledgements}
We are grateful to N. Brambilla, D. Müller, S. Schlichting and N. Tanji and  for discussions. T.~L.\ is supported by the Academy of Finland, project 321840. This work is supported  by the European Research Council, grant ERC-2015-CoG-681707. The content ofthis article does not reflect the official opinion of the Eu-ropean Union and responsibility for the information andviews expressed therein lies entirely with the authors. J.~P.\ acknowledges support for travel by the Jenny and Antti Wihuri Foundation. The authors wish to acknowledge CSC – IT Center for Science, Finland, for computational resources. 
  
\vspace{-0.45cm}

\renewcommand*{\bibfont}{\scriptsize}
\bibliographystyle{elsarticle-num}
\bibliography{spires}







\end{document}